\documentclass[preprint]{aastex}

\begin{document}
\title{A Calibrated Measurement of the Near-IR Continuum Sky Brightness \\ Using Magellan/FIRE}

\author{Peter W. Sullivan\altaffilmark{1}, Robert A. Simcoe\altaffilmark{1}}
\altaffiltext{1}{MIT-Kavli Institute for Astrophysics and Space Research}

\begin{abstract}
We characterize the near-IR sky background from 308 observations with
the FIRE spectrograph at Magellan. A subset of 105 observations
selected to minimize lunar and thermal effects gives a continuous,
median spectrum from 0.83 to 2.5 $\mu$m which we present in electronic
form.  The data are used to characterize the broadband continuum
emission between atmospheric OH features and correlate its properties
with observing conditions such as lunar angle and time of night.  We
find that the moon contributes significantly to the inter-line
continuum in the $Y$ and $J$ bands whereas the observed $H$ band continuum is
dominated by the blended Lorentzian wings of multiple OH line profiles
even at $R=6000$.  Lunar effects may be mitigated in $Y$ and $J$
through careful scheduling of observations, but the most ambitious
near-IR programs will benefit from allocation during dark observing
time if those observations are not limited by read noise.  In $Y$ and $J$ our measured continuum exceeds space-based
average estimates of the Zodiacal light, but it is not readily
identified with known terrestrial foregrounds.  If further
measurements confirm such a fundamental background it would impact
requirements for OH-suppressed instruments operating in this regime.

\end{abstract}

\section{Introduction}

The development of low-noise HgCdTe focal plane arrays has motivated
the construction of a new generation of medium-resolution,
near-infrared spectrometers for ground-based telescopes
\citep{sim08,triplespec,xshooter}.  These instruments resolve the
well-known forest of sky emission line features induced by hydroxyl
(OH) ions and other atmospheric molecules.  For faint object
spectroscopy, instrument sensitivity is therefore limited by the
inter-line sky continuum, which is a superposition of terrestrial,
astronomical, and instrumental backgrounds.

Reliable estimates of the inter-line IR continuum are critical for the
design of successful observations and instruments.  Yet 
calibrated measurements remain sparse in the literature precisely
because the strong foreground line emission makes such a calibration
very challenging.  As we shall demonstrate, the contrast ratio between
narrow emission peaks and the broadband continuum approaches 5 magnitudes in
the $J$ band and 7 magnitudes in $H$.

Attempts to measure the inter-line background accordingly require
certain characteristics of instrumentation and data processing.
First, the spectral resolution must be sufficiently high to separate lines
cleanly.  Second, the detector must exhibit low dark current to not overwhelm the sky signal with thermal shot noise.
Read noise should also be low, but it may be mitigated
by averaging many exposures.  Finally, and perhaps most importantly,
attention must be given to the reduction of scattered light from optical
surfaces in the instrument.  

The FIRE spectrograph at Magellan was designed to address these
criteria while capturing both line and continuum emission of the sky
across the $Y$, $J$, $H$, and $K$ bands simultaneously in each
exposure.  As part of its data processing pipeline, a precise model of
the sky from 0.8 to 2.5 $\mu$m is generated for each science exposure
and saved to disk.

In this paper, we gather these data and present calibrated sky
brightness measurements taken from 308 deep exposures obtained across
several observing runs since the commissioning of FIRE in March 2010.
Section 2 describes the data processing techniques and instrumental
considerations used to arrive at the continuum measurement. In Section
3, we examine correlations between the sky brightness and common
observational conditions, including lunar phase, moon-object angle,
and local time, and we present sky measurements under dark conditions which
minimize their influence. Finally, in Section 4, we discuss the
implications of our findings for the construction of new
instrumentation, planning observations, and allocation of bright
versus dark telescope time for near-IR spectroscopic observations.

\section{Methods}
\subsection{Description of FIRE and FIREHOSE}

The Folded-port Infrared Echellette (FIRE) is a single object,
cross-dispersed spectrometer which delivers $R=6000$ spectra between
$0.82$ and $2.5$ microns in a single setup \citep{sim08}.  It was commissioned on the
Magellan Baade telescope in March 2010 and has been installed
permanently since that time.  

FIRE uses a substrate-thinned HAWAII-2RG HgCdTe detector and SIDECAR
cryogenic readout ASIC to capture the 2-dimensional image of a
$6^{\prime\prime}$-long slit over 21 diffraction orders.  Characterization of the science array's dark current in a laboratory test environment found a
very low count rate of $0.0008\pm0.001$ e$^{-1}s^{-1}$ per pixel or
2.5 electrons per hour.  This is well below the system's read noise and
our measured flux from the sky, so dark current does not contribute to
the absolute sky value nor its measurement uncertainty.

FIRE operates in quasi-Littrow mode and its cross dispersion is
achieved via a network of ZnSe and Infrasil prisms, so the spectral
orders are significantly curved and tilted with respect to the
detector's pixel basis.  We leverage this to obtain sky sampling on
sub-integer pixel scales \citep[similar to the drizzle algorithm
  of][]{drizzle}.  FIRE's optical design retains several arcseconds of
dark pixels between orders to facilitate in estimating and removing
any diffuse scattered light background.  This spacing is largest in
$Y$ and $K$ but somewhat smaller in $H$ because of the balance of the
prisms' partial dispersions.  An additional consequence of the prism
cross disperser is variation in plate scale in the spatial direction
along the slit from order to order.  When estimating surface
brightness, we calibrate the pixel scale for each order empirically by
measuring its width in pixels and scaling to the fixed
$6^{\prime\prime}$ slit length.

FIRE's reduction pipeline (nicknamed FIREHOSE) employs the
2-dimensional sky-subtraction algorithms described by \citet{kel03}
combined with optimal extraction \citep{horne} to produce a
1-dimensional, flux calibrated and telluric-corrected object spectrum.
As part of this process, FIREHOSE creates a 2D estimate of the sky using regions of the slit image not illuminated by object flux. Extracting the sky flux from this image using the same spatial profile as the object results in a 1D sky spectrum. We
have archived these sky estimates from several hundred science
exposures taken with FIRE since its commissioning.  In the raw 2D
frames, one can begin to detect continuum emission between the OH sky
lines, but the single-pixel signal-to-noise ratio can fall below unity in the $Y$ and $J$
bands.  For estimating the continuum flux, we therefore restricted our
analysis to long science exposures with $t_{exp}\geq600$ seconds and
good atmospheric conditions.  The final set of exposures, taken in lunar conditions ranging from new moon to full moon and pointings through 1.00 to 1.86 airmasses, contains 308 frames that constitute 85 hours of integration time spread over 23 nights on 7 separate observing runs.

\subsection{Electronic Offset Correction}

Although the FIREHOSE pipeline applies gain and offset corrections to
the raw 2D images from FIRE, residual offsets remain in the
flat-fielded images (Figure \ref{imgs}a). The value of the offset varies from exposure to exposure; furthermore, it differs for each of the four regions on the H2RG detector which feed separate preamplifier channels in the SIDECAR readout ASIC. The resetting of the preamplifiers, which occurs at the beginning of each frame \citep{sidecar}, is the most likely cause of this offset. Unlike the offset induced by pixel reset, the offset from preamplifier reset is different for each frame and cannot be eliminated through correlated double sampling, Fowler sampling, or sample-up-the-ramp sequences.

During normal data reduction the residual electronic offset is
automatically subtracted from object spectra as part of the sky
estimate.  In order to isolate the sky contribution alone, the
electronic offset must be determined independently from the detector pixels lying between each of the echelle orders on the focal plane. From these pixels, we compute a median offset for the four zones corresponding to each SIDECAR preamplifier, shown in Figure \ref{imgs}b. The typical preamplifier
correction is on the order of 5-10 $e-$, which is comparable to the
sky signal level of 5 to 25 $e-$ (depending on the grating order) as
well as the average read noise for a single pixel, measured from the same non-illuminated pixels to be 7.65 $e-$. However, the large
number of pixels used (at least 167,000 per preamplifier channel) and their fairly even distribution across the detector makes the uncertainty of this correction much lower
than the read noise.

\subsection{Spatially Structured Stray Light Correction}

\begin{figure}
\plotone{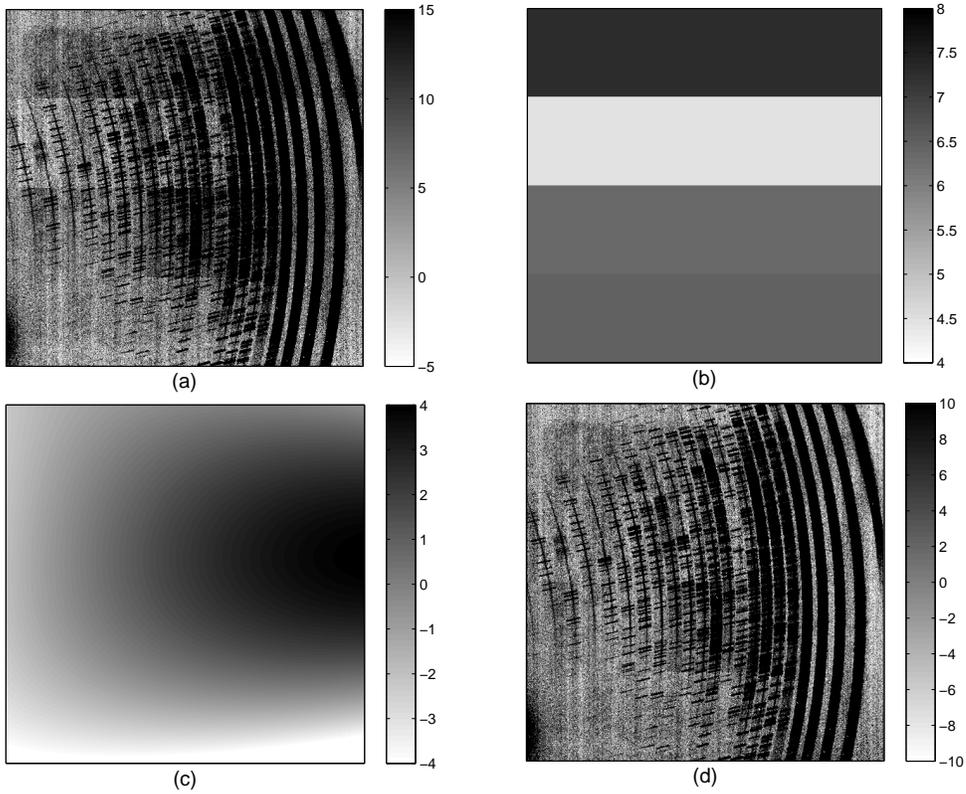} 
\caption{A typical flat-fielded image is
  shown in (a). From the non-illuminated gaps between grating orders,
  we can estimate the offset for each SIDECAR preamplifier (b) and the
  stray-light contribution (c). Subtracting both from the flat-fielded image produces the final result in (d). In practice, the offset and stray light subtraction steps are performed on the 1D sky
  spectrum estimates produced for each echelle order. Note that all four images are in units of  photoelectrons but stretched differently. \label{imgs}}
\end{figure}

After the electronic offset is removed for each SIDECAR preamplifier,
a residual gradient across the detector remains. Because it resembles
a defocused, diffuse version of the echelle illumination, we attribute
it to stray light. The chief source of stray light in FIRE is believed
to be reflections off the front surface of the detector that
subsequently bounce off the facing surface of the last optical element
and return to the focal plane. We can again make use of the pixels lying
between the echelle orders to interpolate the stray light contribution
within the echelle pattern. We fit a third-order, two-dimensional
polynomial to the order-masked pixel intensities to construct
a stray light map for each frame; an example is shown in Figure
\ref{imgs}c. As with the electronic offset, the stray light correction of a few electrons
is of the order of the sky signal itself at this stage. However, 724,000 pixels are used to generate the stray light
estimate, so its uncertainty is negligible relative to that of the sky
signal.

We note that the fitted polynomial falls slightly below zero in the
corners of the array which might seem unphysical. This arises from the
electronic offset correction described in Section 2.2 which over-estimates the
offset in the darkest regions of the array. It is not possible to
separate the true electronic offset from a uniform diffuse
illumination pattern, so we treat the spatially varying portion
of the stray light background simply as a higher-order correction to
any constant offset. The final, corrected image is shown in Figure \ref{imgs}d.

\begin{figure}
\includegraphics[scale=0.8]{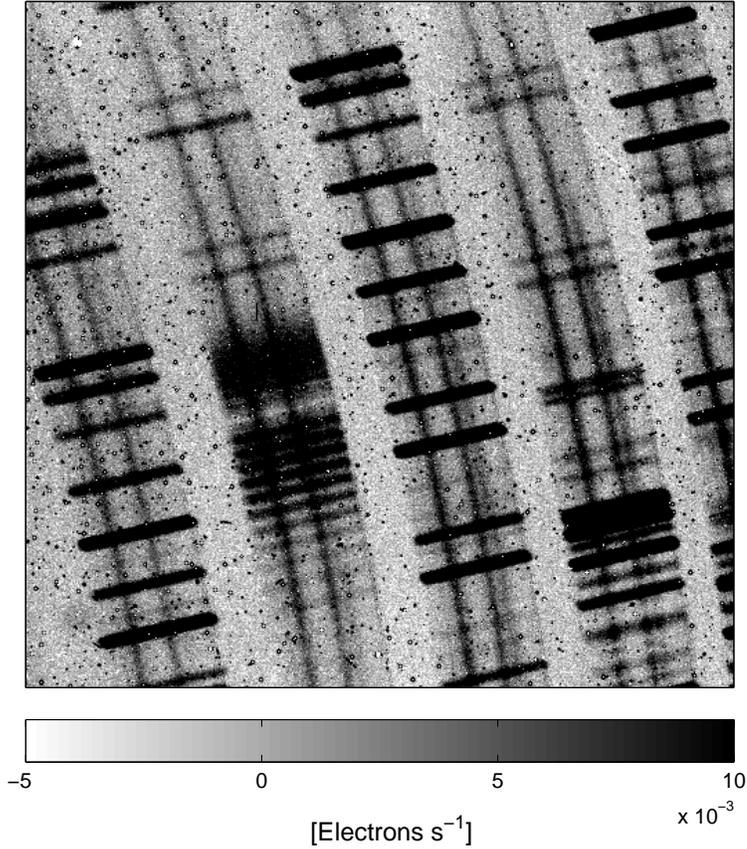} 
\caption{Once the subtractions shown in Figure \ref{imgs} are applied, a stack of 101 images begins to reveal a faint sky continuum in between sky emission lines across the slit. Here, the orders of the echelle image corresponding to the $Y$ and $J$ bands are shown in units of photoelectrons per second. The ratio of sky signal to read noise is approximately 3. The A and B object positions are one- and two-thirds across the slit, where a superposition of object flux is visible perpendicular to the sky lines. The 1D sky spectrum is interpolated at the spatial position of the object flux. Nonuniformities in the slit illumination profile are corrected at a later step in the pipeline, so a slight brightening towards the edge of the slit is visible. The images chosen for the stacking were taken with the moon below the horizon for minimum lunar contribution to the continuum .\label{skyimg}}
\end{figure}

Although the sky continuum is already visible in the $K$ band in
Figure \ref{imgs}d due to thermal emission, the electronic offset and
stray light corrections are required to see the sky continuum in $Y$,
$J$, and $H$. Furthermore, the sky signal from our darkest
observations fall below the single-pixel read noise floor (for one 600 s
exposure) of $Y_{AB}=$18.8, $J_{AB}=$19.2, and $H_{AB}=$19.5 mag
arcsec$^{-2}$. Averaging over many sky spectra or over several
wavelength bins is therefore necessary to reliably measure the sky
continuum.  Figure \ref{skyimg} shows a stack of 101 echelle images
zoomed to the echelle orders corresponding to the $Y$ and $J$ bands
and stretched to reveal the sky continuum above the non-illuminated
gaps between orders.  The stacked images were chosen for their low
lunar contribution (further discussed in Section 3.1), so Figure
\ref{skyimg} represents a worst-case ratio of sky signal to read noise
(approximately 0.3 for a single frame, or 3 for the stack of 101
frames). Because read noise and residual systematics are still much higher than the sky continuum blue-ward of 0.960 $\mu$m, we do not measure the continuum in the $I$ band.

\subsection{Calibration}

\begin{figure}
\plotone{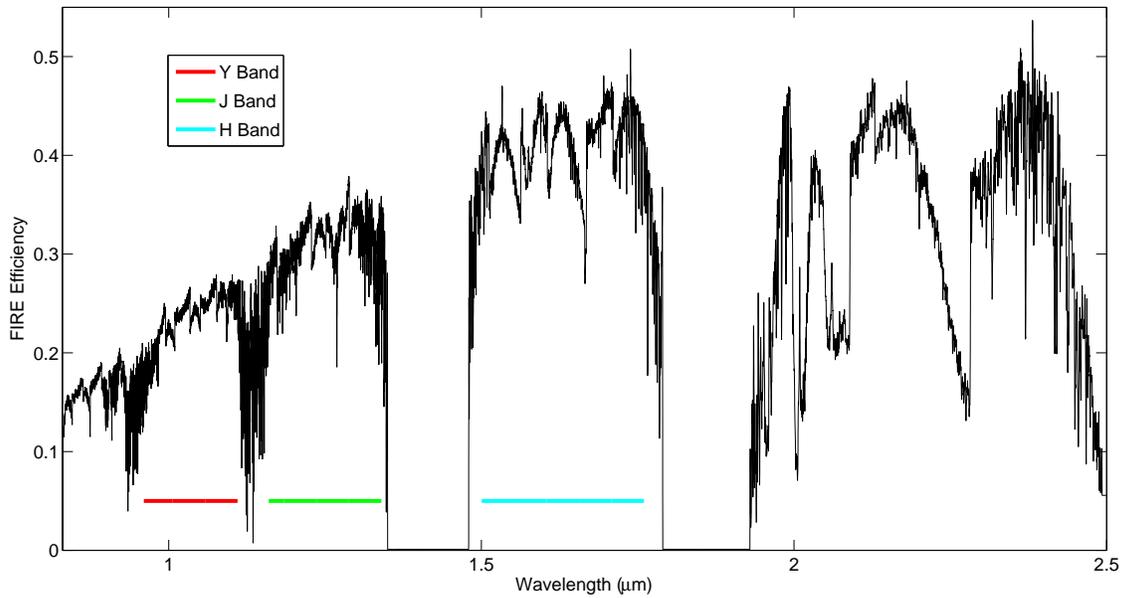} 
\caption{End-to-end efficiency of FIRE determined from comparing
  observations of GD 71 to the CalSTIS archive. Both atmospheric
  transmission and instrument throughput are included in this
  determination. The colored bars show the spectral regions over which
  we measured the continuum sky flux.\label{eff}}
\end{figure}

Flux calibration of our sky frames is performed by comparison with
observations of the hot spectrophotometric standard star GD71 taken
in September 2010.  The optical efficiency of FIRE was determined by
comparing extracted counts to a spectrum taken with HST/NICMOS
available in the CALSPEC
database\footnote{http://www.stsci.edu/hst/observatory/cdbs/calspec.html}. The
resulting efficiency of FIRE as a function of wavelength is shown in
Figure \ref{eff}. This single sensitivity function was used to flux
calibrate all exposures used in this study.

Flux calibration of FIRE science data is typically achieved using
telluric A0V comparison stars rather than white dwarfs, so absolute
spectrophotometric standards were only observed for a small number of
observing runs.  Therefore, we do not have a long term temporal baseline
over which to evaluate FIRE's sensitivity from run to run. 
The GD71 data were taken using a $0.6^{\prime\prime}$ slit and 
experienced slit losses (which do not apply to the sky surface
brightness) since the FWHM of the seeing at the time of the observation was $\sim0.6^{\prime\prime}$.  We performed a crude correction for the slit losses by projecting a circular Gaussian onto the slit profile and estimated the transmitted fraction to be $63\%$.  Given the uncertainty in these parameters, as well as uncertainty in the effective area and
reflectivity of the Magellan telescope mirrors, we conservatively
estimate our systematic uncertainty in absolute calibration of the sky
flux at $\sim 15\%$.

\subsection{FIRE's Line Spread Function}

It has been noted that a significant contribution to the inter-line
background may arise from broad wings of the Line
Spread Function (LSF) which spread light out of the OH lines. Such wings might
originate from scattering at the grating or
other optical surfaces in IR instruments \citep{bland04}.  If
substantial, this effect would reduce the number of pixels that
achieve true background levels. The contribution of the LSF wings to the apparent sky continuum is most important in the $H$ and $K$ bands, where the OH lines are stronger than in the $Y$ and $J$ by 1-2 orders of magnitude.

\begin{figure}
\plotone{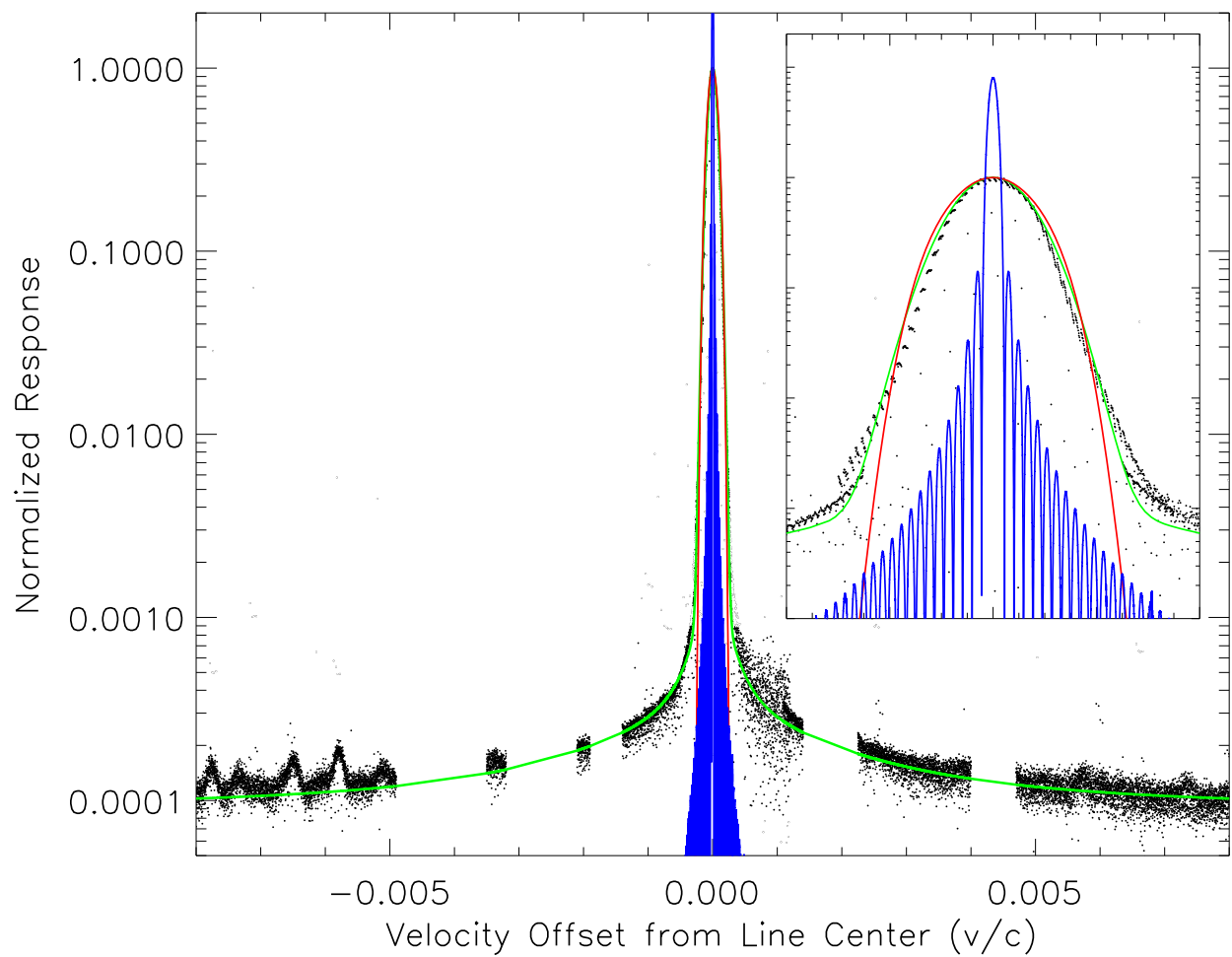} 
\caption{The spectral Line Spread Function for FIRE. Measurement from a ThAr lamp is shown by the black points, and a superposition of Gaussian and Lorentzian functions (green) are fitted. The spectral resolution is limited by the entrance slit, so the grating diffraction pattern (blue) does not dominate the LSF. The total flux of the grating pattern is normalized to the area of the Gaussian core (red).}\label{lsf}
\end{figure}

In Figure \ref{lsf}, we show a high signal-to-noise ratio composite
arc line profile from FIRE where the $x$ axis is scaled to units of
constant velocity $\Delta v/c$ or $\Delta \lambda/\lambda$.  The data
were derived from a deep stack of 20 ThAr arc lamp exposures read at
Fowler-4 to bring out the weak wings of the profile.  The core of the
profile is taken from a bright but unsaturated spectral line while
the wing profile is taken from a nearby line which is saturated in the
core.  The relative normalizations of the core and wings were set by
matching the profile in overlapping regions.  The black dots represent
individual 2D pixel estimates of the flux; several gaps have been
masked to eliminate nearby faint ThAr lines (a handful at $\Delta
v/c<-0.005$ were left in the Figure for reference).

Using these data we explored empirical fits to the LSF functional
form.  First, we verified that the core of the LSF matched
expectations from FIRE's design specifications.  This test is
illustrated with the red curve, which shows a 4-pixel-wide tophat function (the transmission of a 0.6$^{\prime\prime}$ slit) convolved with a Gaussian with a FWHM of 2.0 pixels (FIRE's optical specification), and
further convolved with a kernel describing the MTF degradation from
inter-pixel capacitance in the H2RG detector.  The latter kernel taken from
measurements of the inter-pixel capacitance of FIRE's detector made in a controlled environment, which we model as a leakage of 1.6\% between adjacent pixels.
This core model (Figure \ref{lsf} inset) is an excellent match to the
measured LSF out to $\sim 3\sigma$; the wings of the profile only
begin to contribute when the core has fallen to $0.1-0.2\%$ of its
central value. For reference, the diffraction pattern expected from projecting a uniform circular beam onto FIRE's grating is shown in blue; since the spectral resolution is limited by the slit, it does not dominate the LSF.

At larger separations the wings are much more prominent than the
Gaussian core would indicate; these most likely originate from
scattering off of the gratings and spectrograph optics.
\citet{gratings} studied this effect in detail, fitting the LSF with a
combined Gaussian $+$ Lorentzian $+$ constant offset.  We performed a
similar fit, shown as the green line.  However, FIRE's LSF appears to
have wings on two different spatial scales such that a combination of
two Lorenzian profiles were required: one to fit the shoulder near
$\left|{\Delta \lambda/\lambda}\right| < 0.002$ and another broad component to model power scattered to larger wavelength separations.  In the
later discussion we use this model as our baseline empirical LSF.

\subsection{The Continuum Flux Measurement}
The inter-line sky continuum was computed for each observation separately for the $Y$, $J$, and $H$ bands by finding the mode of the sky flux through an empirical kernel density estimate. The continuum flux levels give a
strong peak in the probability density of the flux (and a well-defined mode), but the sky emission lines are rejected since they can take on virtually any amplitude above the continuum level. Since thousands of spectral bins contribute to the continuum measurement, the low signal-to-noise ratio for a single pixel is easily improved by this approach. Only in the $K$ band is the continuum level ill-defined because thermal emission from the telescope and atmosphere causes the background to rise sharply with wavelength.  Since these effects are site-specific and do not reflect the fundamental sky background, we do not compute the $K$ sky brightness.

\section{Data and Discussion}

\subsection{Effects of Moon, Observation Time, OH Flux}
\begin{figure}
\plotone{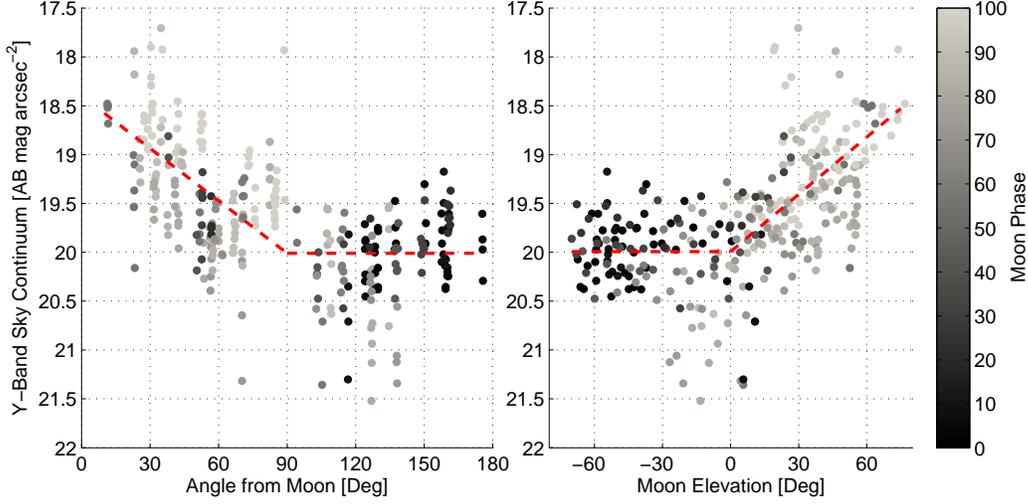} 
\caption{In the $Y$ band, angular distance from the moon and moon
  elevation impact the sky continuum most strongly. The linear
  least-squares fits were only performed on observations with the moon above the horizon and $>50\%$
  illumination. The mean dark level is plotted for negative moon elevations. The slope of the fits are shown in Table \ref{tbl}. \label{ymoon}}
\end{figure}

Since the sky continuum has been quantified under a variety of
observing conditions, it is possible to determine how the dark sky
levels quantified in Section 2.6 change with the influence of moon,
local time, and the OH line flux. It is well-established that lunar
illumination increases the inter-line sky continuum in the optical; the
short-wave end of FIRE overlaps with this region and may suffer from
lunar sky brightening as well. As shown in Figure \ref{ymoon}, the sky
continuum levels measured in the $Y$ band do indeed decrease with
angular separation from the moon and increase with moon elevation. The
horizontal portions of the fits show the mean sky level under moonless
conditions, and the sloping portions show a linear least-squares fit
to the sky magnitude from the observations where the moon is
present. The slopes computed in all three bands are summarized Table
\ref{tbl}. The sky continuum is most strongly correlated with moon in
the $Y$ and uncorrelated in $H$. Due to the desire to observe at low
airmass, observations taken with a high lunar elevation largely
coincide with observations having a small moon-object angle. This
degeneracy makes it difficult to determine which factor has a greater
impact upon the sky continuum in the $Y$ and $J$.

\begin{figure}
\plotone{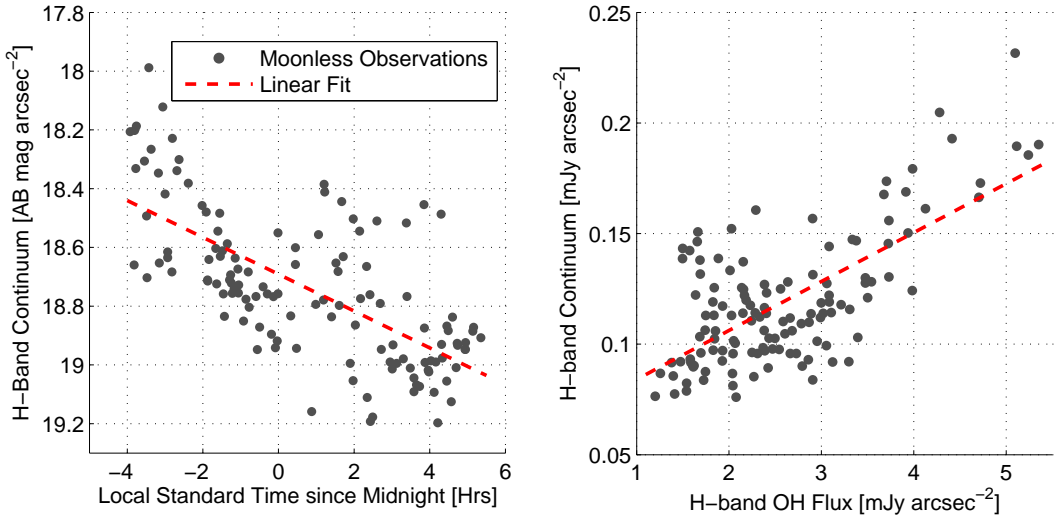} 
\caption{In the $H$ band, the local time of night (left) has the largest correlation with continuum flux as the telescope and its surroundings cool down. The OH flux (right) shows a strong correlation with the continuum measurement as the wings of the LSF allow flux from the OH lines to scatter into the continuum. The linear least-squares fits were performed only on the moonless observations, and the slopes are shown in Table \ref{tbl}. \label{htime}}
\end{figure}

At longer wavelengths, thermal emission from the telescope,
instrument, and atmosphere decreases as the night progresses. In
addition, the OH line strength is strongest in the evening twilight and
can induce a higher measured continuum via the wings of the
LSF. Thermal and OH flux dependencies therefore trend with the local
time, which indeed is moderately correlated with sky continuum in the
$H$ band. In Figure \ref{htime}, a linear fit to local time has been
performed on the observations having a moon angle of more than 90
degrees and presented in Table \ref{tbl}. The OH line flux can be separately measured as the integrated
flux in the band above twice the continuum level. The correlation between
the $H$ continuum and OH flux, as shown in Table \ref{tbl}, is quite strong, suggesting that the Lorentzian wings of the LSF
may account for the observed inter-line continuum in $H$. Due to the noticeable effects of thermal and OH emission, we do not find observations taken before 2200 local standard time to be ``dark''.

As summarized in Table \ref{tbl}, the $Y$ band continuum shows no
significant dependence on the local time or OH flux, and the $H$ band
continuum shows weak dependence on lunar effects. Furthermore, the sky
continuum shows no clear correlations with ecliptic declination,
airmass, or solar angle in any of the bands.

\subsection{Dark Sky Measurements}
\begin{figure}
\includegraphics[scale=0.8]{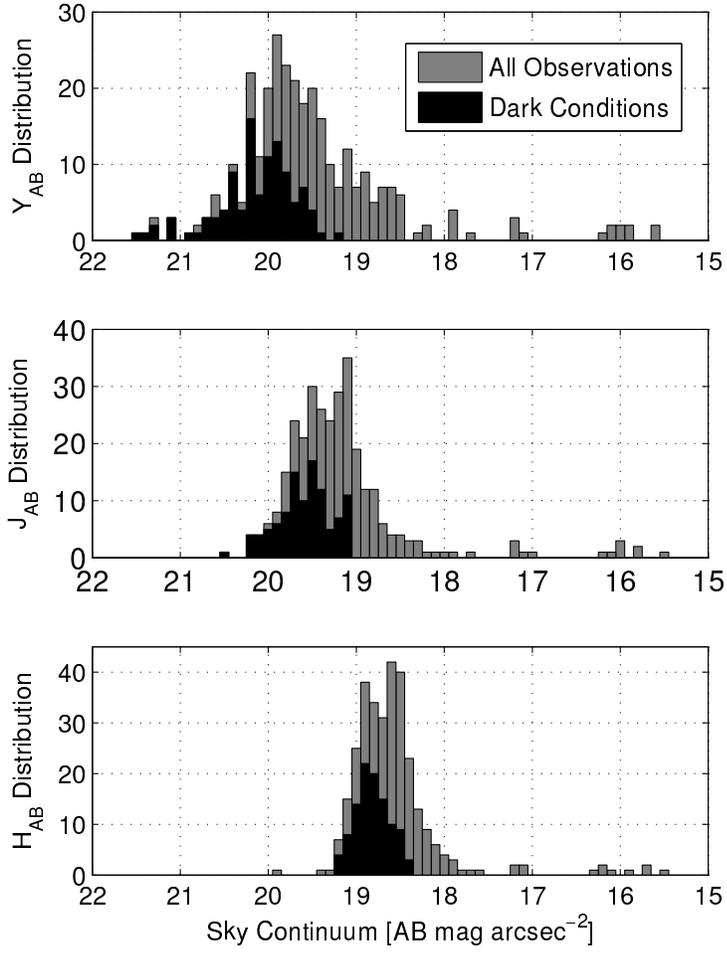} 
\caption{Continuum sky background in the $Y$, $J$, and $H$ bands. The
  specific spectral regions used for each determination are shown in
  Figure \ref{eff}.
\label{hists}}
\end{figure}

The distribution of measured continuum values for each band are shown
in Figure \ref{hists}. To quantify the true sky continuum level under
dark conditions, we selected the sky measurements with minimal
contamination from the moon and thermal emission. As justified in the
previous section, we define ``dark'' observations to be those taken
with a moon-object angle of at least 90 degrees and no earlier than
two hours before local midnight. There were 105 observations which met
these criteria (constituting 31 hours of integration time), and the
distributions of the dark subsets are also shown in Figure
\ref{hists}.
 
For the dark observations, we measure mean inter-line sky continuum levels of $Y_{AB}=$20.05 $\pm0.04$, $J_{AB}=$19.55 $\pm0.03$, and $H_{AB}=$18.80 $\pm0.02$ mag arcsec$^{-2}$. Besides the statistical
uncertainties listed here, each measurement has a systematic
uncertainty of +0.15/-0.18 AB mag arcsec$^{-2}$ that stems from the
efficiency calculation described above. We also computed the median
sky spectrum that is available in electronic form. The sky spectrum
has read noise floors of $Y_{AB}=21.35$, $J_{AB}=21.75$, and
$H_{AB}=22.07$, which is below the continuum levels by more than one
magnitude for wavelengths greater than 0.960 $\mu$m.

\subsection{Comparison with Previous Sky Estimates}

When developing designs for FIRE, we relied on two estimates of the sky continuum: \citet{mai93} and the Gemini Observatory sky
model\footnote{http://www.gemini.edu/?q=node/10787}.  The Gemini
model was generated using a combination of ATRAN \citep{atran} models
and the theoretical OH emission spectrum of \citet{rousselot}. Since
FIRE's construction, the Gemini model has been revised, so we compare
to the latest version (uploaded 16 April 2010). We also compare to the measurements of \citet{cuby_messenger} and \citet{ellis2012}.

\begin{figure}
\includegraphics[scale=0.7]{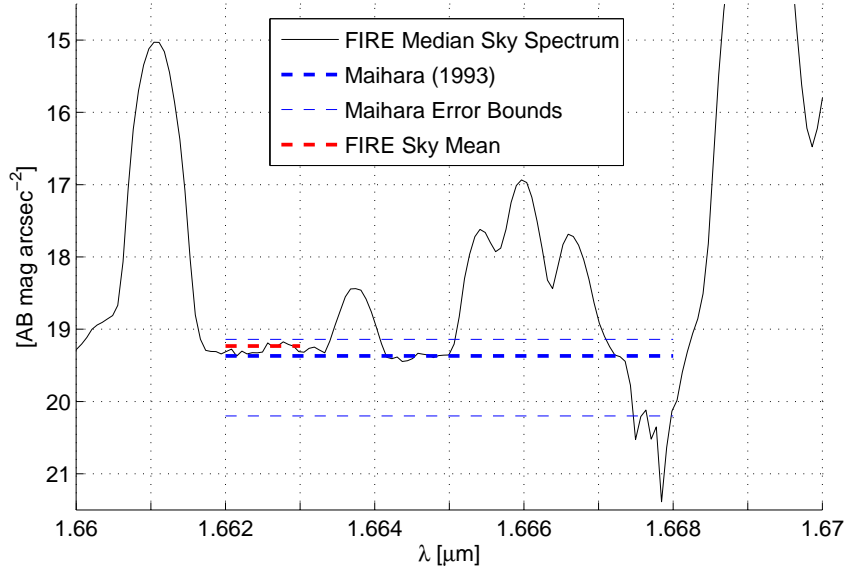} 
\caption{The mean sky spectrum over the same region shown in Figure
  2 of \citet{mai93}. The blue line indicates the sky level measured by
  Maihara at Mauna Kea, and the red bar indicates the region over
  which our median sky spectrum is compared to it. The low-level
  emission and absorption features ($\pm2$ mag) are likely due to atmospheric
  methane.\label{meth}}
\end{figure}

Maihara's measurement, made with the UH 2.2 meter telescope on Mauna
Kea, reports a flux of 590 ($\pm$140) $\gamma$ s$^{-1}$ m$^{-2}$ $\mu$m$^{-1}$ arcsec$^{-2}$ or 19.4 ($\pm$0.3) AB mag arcsec$^{-2}$ over
the range of 1.661 to 1.669 $\mu$m. A direct comparison to our
measurement is complicated by low-level emission features near 1.665
$\mu$m (attributed to atmospheric methane) that are not apparent in
the Mauna Kea data. Considering only the range of 1.662 to 1.663
$\mu$m (as shown in Figure \ref{meth}) gives a sky flux of 670 ($\pm$200) $\gamma$ s$^{-1}$ m$^{-2}$ $\mu$m$^{-1}$ arcsec$^{-2}$ or 19.2 ($\pm0.2$) AB mag arcsec$^{-2}$ measured by FIRE, which is well within
the upper bound of Maihara's uncertainty. Our measurement across the entire $H$ band of 18.8 ($\pm$0.2) AB mag arcsec$^{-2}$ is also consistent with \citet{ellis2012}, who measures 860 ($\pm$210) $\gamma$ s$^{-1}$ m$^{-2}$ $\mu$m$^{-1}$ arcsec$^{-2}$ or 19.0 ($\pm0.3$) AB mag arcsec$^{-2}$ in the $H$ band.

The Gemini sky model shows a darker continuum flux of 19.63 AB mag
arcsec$^{-2}$ at 1.663 $\mu$m. However, at 1.25 $\mu$m, the Gemini
model gives 19.35 versus our measurement of 19.55 AB mag arcsec$^{-2}$ across $J$, and at
1.02 $\mu$m, the model gives 19.23 versus our measurement of 20.05 AB mag
arcsec$^{-2}$ across $Y$. Maihara's measurements do not cover the $J$ or $Y$
bands, but provisional measurements of the inter-line continuum in $H$ and $J$ were
made during commissioning of the ISAAC spectrometer on the ESO/VLT 
\citep{cuby_messenger}.  The ISAAC team reports backgrounds of $17.88$ in $H$ and $18.94$ AB mag arcsec$^{-2}$ in $J$; in linear units this is 2.3 times higher than our measurement in $H$ and $\sim75\%$ higher than our measurement in $J$.  One might expect a significantly higher background at the lower and warmer site of the Magellan Telescopes at Las Campanas relative to Mauna Kea or Paranal, but we find this is not the case.

\subsection{What Fraction of the Sky is OH-Free?}

For the design of IR spectrometers, it is vital to understand what
fraction of the infrared sky is ``clean'' of OH contamination and how
this quantity varies with spectral resolution.  This is particularly
true in light of the broad wings present in the LSF for FIRE (or
indeed any known spectrometer).  The above analysis indicates that
some OH flux enters the continuum (especially in the $H$), but it is
possible that {\em all} of the broadband flux we have attributed to
the continuum is in fact a superposition of wings from many
neighboring OH features \citep{OHSuppression}.

With a working model of the LSF, we may test this hypothesis by
constructing a simulated, high-resolution sky containing only OH lines,
convolving it with the measured LSF and comparing its resultant
inter-line continuum with our measured values.  To this end, we used
the OH line lists of \citet{rousselot} to set the wavelengths and
relative intensities of the OH forest lines in a high resolution
($R=30,000$) spectrum consisting of simple Gaussian profiles.  All
line peak heights were normalized to the data using a single, constant
scale factor, which was determined by enforcing that the average total
flux in the model lines equal the average total flux in our line
measurements for the true sky.  For the high resolution spectrum this
corresponds to narrower and taller peaks but with the same integrated
area.  We then convolve with the LSF determined above, finding (as
expected) that the height and width of the individual lines match in
detail.

\begin{figure}
\plotone{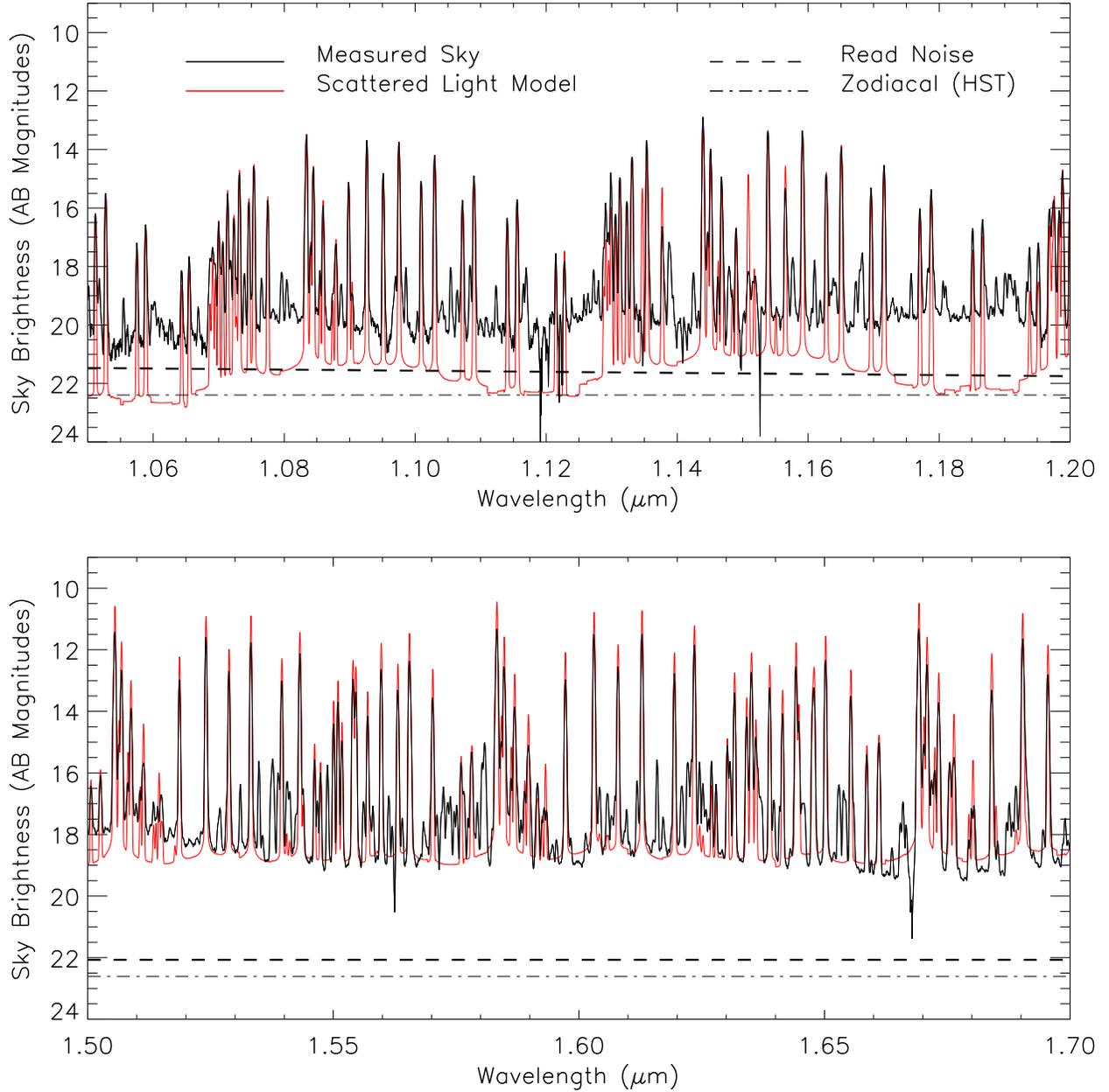} 
\caption{In the $H$ band (lower panel), the convolution of our empirically-determined LSF with an OH linelist (red) shows good agreement with the measured sky spectrum (black). Across the $Y$ and $J$ bands (upper panel), the measured sky spectrum shows a higher continuum level than the LSF convolution would predict. In all three bands, the measured continuum level is well above the read noise floor and Zodiacal light.}\label{ohscatt}
\end{figure}

Figure \ref{ohscatt} illustrates the result of this calculation,
where the red curve shows the model OH forest convolved with the
LSF and the black shows our measured median sky spectrum. We find that
in $H$, scattered light from the wings of the LSF can indeed account
for the full measured background flux seen in FIRE, so the two curves
agree quite well. However, in $Y$ and $J$ the convolved curves fall
well below our measured background, indicating that the continuum in
these bands arises from a source other than the OH forest. The read noise floor of our median sky spectrum and HST zodiacal light levels are also plotted, but they fall far below our measured spectrum.

This dichotomy may be easily understood in terms of dynamic range and
line density.  The LSF presented in Figure \ref{lsf} clearly shows
that for intrinsic line-to-continuum ratios smaller than five magnitudes, the
inter-line spectral regions never ``see'' the LSF wings, so only the core
emerges from the background.  At line-to-continuum ratios $\gtrsim 5$ mag, the wings become more prominent and can extend for many
pixels.  In $Y$ and $J$, our observed line-to-continuum ratio of 2-5 mag puts the blended OH background well below the continuum value.  Only in $H$ are we in
the region where OH blending dominates.

\begin{figure}
\includegraphics[scale=0.7]{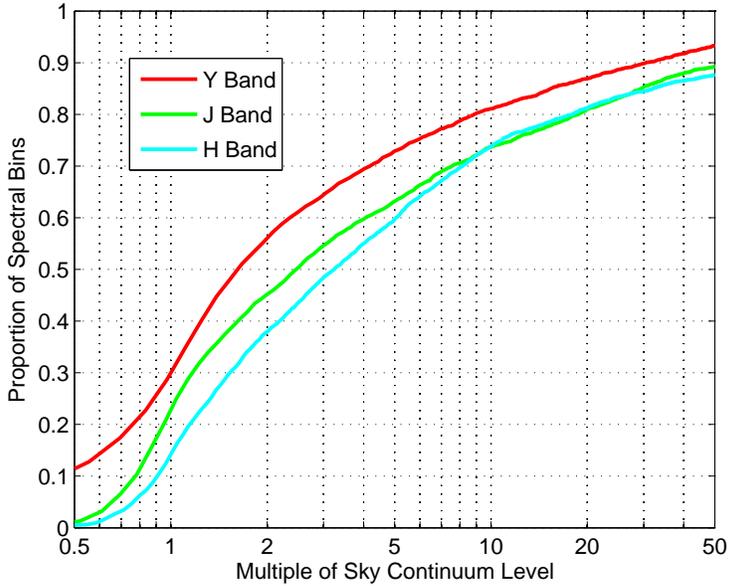} 
\caption{The cumulative proportion of spectral bins achieving the sky continuum level, and multiples thereof, are plotted for the Y, J, and H bands. The specific spectral regions used for each determination are shown in Figure \ref{eff}.\label{cdf}}
\end{figure}

To address the question heading this section, it appears that there is
no truly OH-free spectrum in FIRE's $H$ band channel as blending of
the LSF's wings limits the background, commensurate with the findings
of \citet{OHSuppression}. Our extrapolation to zero OH flux (Table
\ref{tbl}) suggests that a continuum of $H_{AB}=19.4$ AB mag
arcsec$^{-2}$ may be achievable. Blue-ward of $H$, a fundamental
background is reached by significant proportions of the spectrum. To
better quantify this statement, we plot in Figure \ref{cdf} the
cumulative sky flux distribution renormalized so that a value of unity
corresponds to the continuum level over each bandpass. In $Y$,
approximately $30\%$ of the spectrum is at the background level and
$75\%$ of the spectrum is within a factor of 6 of the background.  In
$J$ and $H$, less than $25\%$ is at the background, and the 75\% mark
is only reached at 10 times the background.

To quantify the effects of spectral resolution on the $Y$ and $J$
measurements, we recalculated the cumulative sky flux distribution for
degraded instrument resolutions. FIRE's resolution is slit-limited so
one may convolve with tophat kernels of varying width to simulate the
effect of changing slits.  However, in this analysis we wished to
simulate instruments of varying spectral resolution that are assumed
to critically sample a LSF at a specified resolution, so a Gaussian
kernel is more appropriate. There is no need to include the Lorentzian
wings in the convolution since the continuum level dominates in these
bands.

\begin{figure}
\includegraphics[scale=0.7]{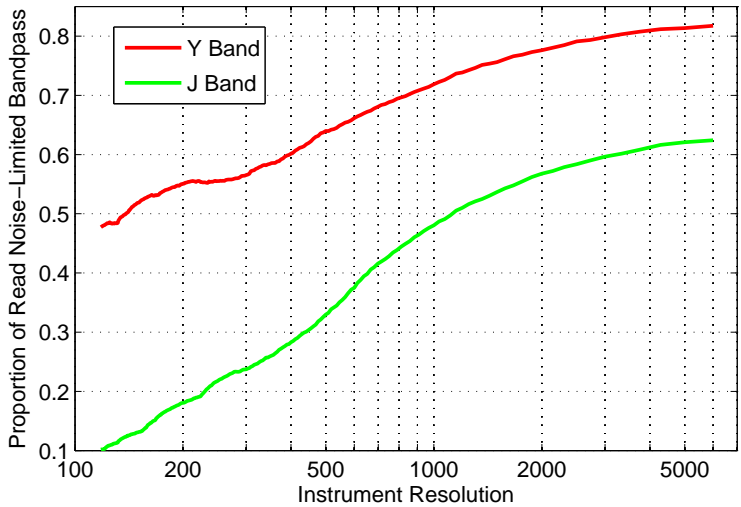} 
\caption{The proportion of spectral bins having sky noise lower than the read noise of 7.65 electrons for an exposure of 1500 seconds are plotted against instrument resolution. The data points at $R=6000$ are from FIRE's sky spectrum; all lower resolutions were simulated by convolving the FIRE spectrum with Gaussian kernels of increasing width.}\label{res}
\end{figure}

Following \citet{martini}, we plot in Figure \ref{res} the proportion
of spectral bins which are read-noise dominated versus resolution. We
see that as $R$ increases from 500 to 3000, a critical change occurs
as continuum begins to emerge between OH peaks. Martini assumed a
lower value of the detector read noise (5 e- versus our operational
value of 7.65 e-), so we compensate with a longer effective exposure
time (1500 s versus 600 s) in order to make the plots comparable. In
the $J$ band, we arrive at essentially the same proportion of read
noise-dominated bins at R=6000 as Martini, but at R=500, we predict
that 33\% of spectral bins are read noise-dominated where Martini
predicts 0\%. In the $Y$ band, we show that at R=6000, 83\% of
spectral bins are read noise-dominated, and at R=5000, 64\%.

Since the spectral bins at $R=6000$ that reach the continuum level are read noise-dominated, one can alternatively determine the exposure time required to reach sky-dominated noise. This is a critical limit to calculate for any instrument observing faint objects. Assuming a sky background of 20.05 AB mag arcsec$^{-2}$ in $Y$, FIRE receives a photon count from
the sky of
\begin{equation}
f = 0.051 ~\gamma ~{\rm sec}^{-1} \left( {{D_{tel}}\over{6.5 {\rm
      m}}}\right)^2 \left({{\Delta v_{pix}}\over{12.5 ~{\rm
      km/s}}}\right)^{-1}\left({{w_{slit}}\over{0.6^{\prime\prime}}}\right)\left({{h_{pix}}\over{0.125^{\prime\prime}}}\right)
\end{equation}
where we have used scalings for the telescope aperture $D_{tel}$, the
velocity interval per pixel $\Delta v_{pix}$, the slit width
$w_{slit}$ and pixel scale $h_{pix}$ in the spatial direction
appropriate for Magellan/FIRE.  To translate to a noise budget, we
assume $\sim 24\%$ instrument efficiency in $Y$ and $J$ (Figure \ref{eff})
and a read noise RMS of $7.65e^-$, appropriate for Fowler-8 or long
Sample-Up-The-Ramp integrations.  For these typical conditions,
detector read noise dominates the total budget for exposures shorter
than $t_{exp}\approx 4800$ seconds, or 80 minutes. Our original designs for FIRE were based on a slightly brighter estimate of the inter-line continuum, which would have delivered sky-dominated exposures in $\lesssim 10$ minutes. Hence, this minimum time to sky-dominated noise is longer than we consider optimal.  

\subsection{Sources of Inter-Line Continuum Emission in the $Y$ and $J$ bands}

The inter-line continuum in $Y$ and $J$ substantially
exceeds our scattered light models derived from the empirical LSF by 1-2
magnitudes or a factor of 2 to 6.  This suggests the existence of
another fundamental diffuse background for terrestrial observations,
which would have implications for the efficacy of dark and/or
OH-suppressed imaging and spectroscopy in this regime.  However, this
detection remains puzzling as there is not a readily identifiable
emission mechanism in the Earth's atmosphere that would give rise to
such a background.

We were initially concerned that our ``measurement'' might actually reflect a
read noise floor rather than a statistically significant detection.
However, Figure \ref{skyimg} illustrates an enhancement in the diffuse
flux between adjacent OH lines relative to the non-illuminated gaps
between echelle orders even under dark conditions. Furthermore, in Figure \ref{ohscatt}, the
effective surface brightness of the read noise (defined as the
magnitude corresponding to a $1\sigma$ constant offset from zero flux) is plotted with the sky measurement. 
Comparison of the read noise floor and the measurement curve
clearly shows detection at a signal-to-noise ratio varying with
wavelength from $\sim2.3\sigma$ to $\sim6\sigma$.

Beyond instrumental effects, reflected Zodiacal light (ZL) is considered the
fundamental continuum limit in the near-IR once the OH forest has
been mitigated.  This background has been measured at $1.2 \mu$m from
orbit using HST/NICMOS with an average value of 22.4 AB
mag arcsec$^{-2}$ \citep{zodi_nicmos}.  This is 2.8
magnitudes (or a factor of $\sim14$) fainter than our measurement at $J$.

In addition to instrument-scattered OH, read noise, and the Zodical
backgrounds, other possible sources of diffuse light include
unresolved stellar populations (primarily Red Giants in the near-IR) as well as tropospheric scattering of all the above backgrounds combined with terrestrial light pollution \citep{leinert}.  We expect the
contribution from unresolved stars with $J>20$ to be minimized since
most of the data used to determine the stack were obtained from
observations of high-redshift quasars, which are located
preferentially out of the Galactic plane.  At high galactic latitude $b$, the integrated
1.25 $\mu$m stellar background estimated from DIRBE is 10-100 times
lower than in the plane and should in fact fall somewhat below the
average ZL background \citep{leinert}. Even the resolved
stellar background will have a small fraction of radiation scattered by the troposphere into a diffuse component, but it too is mitigated at high $b$.  Atmospheric turbulence and scattering of the very
strong OH forest does not distort the spectrum (at the
resolutions considered here) and should not spread on-resonance OH
emission into the dark inter-line regions.

Light pollution is also expected to play a minimal role as Las
Campanas is a substantially dark site and the stacks presented here
were selected for dark lunar conditions.  Presently, measurements do
not exist for the level of light pollution at LCO, but corresponding
studies at CTIO/Cerro
Pachon\footnote{http://www.ctio.noao.edu/site/pachonsky/} ---both of
which are closer to the cities of La Serena/Coquimbo---remain quite
dark and the contribution of light pollution falls below that of the
ZL over most of the sky.

It appears that a robust measurement of the $Y$ and $J$ backgrounds will
remain challenging for some time to come, and further estimates of
this background with other instruments of varying resolution will be
important for verifying the results reported here.  Despite our
careful attempts to correct for diffuse and grating-scattered light in
our optical system, there is a faint signal detected which cannot be
identified with a simple atmospheric emission mechanism.  Since this
emission is only detected at $\sim 2-5\sigma$ significance in a $>20$
hour stacked spectrum on a 6.5 meter telescope, we may be susceptible
to systematic effects beyond those considered here.  Since we are
read noise-dominated in this regime for single exposures, it will be
particularly important to test this result against spectra obtained at
lower resolution in low-background instruments with HAWAII-2RG
detectors such as the newly-commissioned MOSFIRE spectrograph
\citep{mosfire}.  These may be more sensitive to blended wings of the
OH lines but could provide a crucial check of the broadband signal in
relatively un-blended regions.

\section{Implications for Telescope Allocation and Instrument Design}

Traditionally, infrared spectrometers (and imagers) are scheduled for
use during the bright phase of the lunar cycle under the premise that
the OH sky is of comparable or greater brightness than the moon and
therefore dominates the background.  This is clearly the case for imagers
and low-resolution spectrometers, but with low-background, moderate-resolution instruments, we must consider whether there is a regime where the moon makes a substantial contribution to the noise
background.  Such a finding would have implications for best observing
practices and affect whether telescope allocation committees
should consider proposals to perform IR spectroscopy during coveted
dark time.

Our simple regression and OH blending analyses indicate that lunar
illumination has a negligible impact in the $H$ and $K$ bands but does matter
in $Y$ and $J$.  Our data show that when the moon is set
and/or dark, the background converges to its lowest values, between
$Y_{AB}=19.5$ and 20.5.  When the moon is nearby or
overhead, the background is over a magnitude ($2.5\times$) brighter.
This amounts to a $58\%$ increase in noise over the dark-sky case,
{\em if} the instrument is background-limited.  Put differently, the
same $Y$-band observation will achieve, on average, a 58\% improvement
in signal-to-noise ratio when undertaken in dark skies, which is
equivalent to increasing the telescope's primary aperture by the same
factor.

We therefore conclude that the sky is sufficiently dark in $Y$ and $J$
that the most aggressive science programs, which are sky-noise
dominated, would indeed benefit from scheduling during gray or dark
time.  Some portion of the benefits may be obtained by careful
scheduling of targets to maximize moon angle, ideally to
$\gtrsim 60$ degrees.  During the full moon this becomes
impractical when coupled with the desire to observe at low airmass, so
some compromise must be reached.  Science programs relying primarily
on $H$ or $K$ may be observed equally well during bright or dark time.  

The low sky background and associated tradeoff with read noise have
informed our observing and data reduction strategies for FIRE (particularly for high-redshift targets.)  Since the exposure times
required to achieve sky dominated noise are significantly longer than
the coherence time of the OH sky lines, standard pairwise image
subtraction techniques (e.g. ``ABBA'' nods) do not perform well for
sky subtraction.  Instead, we now acquire the faintest targets in long
sample-up-the-ramp integrations ($\sim 30$ minutes) and reduce each
frame using a sky model generated on the fly.  This approach is time
consuming and computationally expensive, but it has the dual advantages
of mitigating read noise and increasing the signal-to-noise ratio relative to pairwise subtraction by a factor of $\sim\sqrt{2}$.

Our sky measurements may warrant consideration in the design
of future IR spectrometers for large aperture telescopes.  As one
moves to high spectral resolution or narrow-slit, AO-fed
instrumentation, the benefits for faint object work may not be fully
realized unless telescope aperture keeps pace with the associated
reduction in sky background or if there is a concerted effort to battle
sensor read noise to even lower values.  Both approaches require
considerable resources and careful trade studies would be needed to
optimize allocation between the two.  

If the inter-line background measured in the $Y$ and $J$ is supported by
further observations, then it would imply that ground-based
observations could only reach within $\sim 3$ magnitudes of the
Zodiacal background in the near-IR, or slightly more than a factor of 10.
This background would slightly weaken the case for OH-suppressed instruments \citep[e.g., ][]{ellis2012}, but it would also
lessen the technical requirements for OH-suppressed instruments to
achieve background-limited performance in $J$ from $\sim 30-35$ dB
narrowband attenuation to $\sim 20$ dB. These arguments are particularly
germane for studies of the Lyman alpha transition in high-redshift
objects since it is observable in the dark portions of $Y$ and $J$ bands throughout
the epoch of cosmic reionization.

\section{Conclusions}

We have presented a composite spectrum of the near-infrared night sky
obtained by stacking 308 exposures obtained with Magellan/FIRE on 23
nights over seven observing runs.  Careful attention was given to
correcting electronic artifacts from the detector and scattered light
in the instrument.  Absolute flux calibration was taken from hot
spectrophotometric standards, although the systematic error in the 
instrument efficiency is uncertain at the $\sim 15\%$ level because of
slit losses. Analysis of a high signal-to-noise arc lamp composite
indicates a small amount of OH line flux is scattered into broad
Lorentzian wings at the level of $\sim 10^{-4}$ times the peak line
intensity. We explored measurements of FIRE's inter-line sky continuum
and compared to previous estimates with the following main results:

\begin{enumerate}
\item{The mean of the inter-line continuum falls at
  $Y_{AB}=20.05\pm0.04$, $J_{AB}=19.55\pm0.03$, and
  $H_{AB}=18.80\pm0.02$ (stat.) $\pm 0.2$ (sys.) mag
  arcsec$^{-2}$. This is consistent with what is reported in
  \citet{mai93} for the $H$ band.}
\item{The $H$ band background correlates most strongly with OH
  intensity since the $H$ band ``continuum'' is largely a
  superposition of scattered light from OH emission features. OH intensity decreases
  after sunset, giving a moderate correlation between $H$ continuum flux and time of night, but thermal emission also contributes. Our $H$ band measurement and that of
  \citet{mai93} may not achieve the true $H$ continuum, which we
  extrapolate to being 19.4 AB mag arcsec$^{-2}$.}
 \item{The $Y$ and $J$ band inter-line continuum fluxes correlate with
    moon-object angle and moon elevation, which are degenerate in our
    data set.  Observations obtained at $\lesssim 30$ degrees from a
    gibbous to full moon exhibit backgrounds 2.5 times brighter on
    average than those from a dark sky.  This trend is much weaker in
    $H$ and not seen in $K$.}
\item{Under dark conditions, the $Y$ and $J$ continua are still higher than the
  predicted signal from scattered OH emission, and our measurements
  seem to detect a true broadband background.  At $R\sim 6000$, 60-80\%
  of spectral bins achieve this background level.  If real, it exceeds the average level of Zodiacal light measured by
  HST/NICMOS, yet it is not easily explained by known terrestrial
  foregrounds.  }
\item{The most challenging programs in $Y$ and $J$ will benefit from
  obtaining telescope allocations during the dark half of the lunar
  cycle {\em provided} that instruments are sufficiently sensitive
  (or exposures are sufficiently long) for sky noise to dominate the
  higher read noise of CMOS/HgCdTe detectors over CCDs.}
\end{enumerate}

\acknowledgements

The authors wish to thank the staff of the Magellan Telescopes and Las
Campanas Observatory for their assistance in obtaining the
observations described herein.  The majority of the data were taken
for programs supported under NSF Grants AST-0908920 and AST-1109115.
RAS thanks D. Schelegel, S. Perlmutter, and D. Fabricant for providing
continued motivation to see this analysis through. RAS was supported
during this research by the Adam J. Burgasser Chair in Astrophysics.

\bibliography{ms}

\clearpage

\begin{table}
\begin{center}
\caption{Correlations between sky continuum and lunar, thermal, and hydroxyl effects.\label{tbl}}
\begin{tabular}{rccc|ccc}
\tableline\tableline
 & \multicolumn{3}{c}{Fit to Moon Angle} & \multicolumn{3}{c}{Fit to Moon Elevation} \\
 & Value\tablenotemark{a} at 90$^{\circ}$ & Slope\tablenotemark{b} & $R^{2}$ 
 & Value\tablenotemark{a} at 0$^{\circ}$  & Slope\tablenotemark{b} & $R^{2}$ \\
\tableline
$Y$  & 20.01 & 17.9 & 0.273 & 19.99 & -19.5 & 0.194\\
$J$  & 19.50 & 11.2 & 0.245 & 19.47 & -10.7 & 0.110\\
$H$  & 18.71 & 4.11 & 0.093 & 18.68 & -2.50 & 0.010\\
\tableline
& \multicolumn{3}{c}{Fit to Local Time} & \multicolumn{3}{c}{Fit to OH Flux}\\
& Value\tablenotemark{a} at Midnight & Slope\tablenotemark{c} & $R^{2}$ 
& Value\tablenotemark{d} at 0 Flux   & Slope\tablenotemark{e} & $R^{2}$\\
\tableline
$Y$  & 20.05 & 56.0 & 0.119 & 0.0312 & 0.013 & 0.025\\
$J$  & 19.50 & 54.3 & 0.237 & 0.0543 & 0.003 & 0.044\\
$H$  & 18.69 & 62.7 & 0.461 & 0.0619 & 0.022 & 0.462\\
\end{tabular}
\tablenotetext{a}{AB mag arcsec$^{-2}$.}
\tablenotetext{b}{AB mmag arcsec$^{-2}$ deg$^{-1}$.}
\tablenotetext{c}{AB mmag arcsec$^{-2}$ hr$^{-1}$.}
\tablenotetext{d}{mJy arcsec$^{-2}$.}
\tablenotetext{e}{Dimensionless in linear units, e.g. mJy arcsec$^{-2}$ per mJy arcsec$^{-2}$.}
\end{center}
\end{table}

\end{document}